%Paper: hep-th/9510103
%From: Konstadinos Sfetsos <K.Sfetsos@fys.ruu.nl>
%Date: Mon, 16 Oct 1995 13:12:41 +0100 (GMT+0100)
%Date (revised): Mon, 16 Oct 1995 15:22:20 +0100 (GMT+0100)

%%%%%%%%%%%%%%%%%%%%%%%%%%%%%%%%%%%%%%%%%%%%%%%%%%%%%%%%%%%%%%%%

\input harvmac

%%%%%%%%%%%%%%%%%%%%%%%

%\draftmode
\noblackbox
\baselineskip 14pt plus .1pt minus .1pt

\overfullrule=0pt

%%%%%%%%%%%%%%%%%%%%%%%%%%%%%%%%%%%%%%%%%%%%%%%%%%%%%%%%%%%%%%%%

% definitions

\def\bs{\bigskip}
\def\no{\noindent}
\def\hb{\hfill\break}
\def\qq{\qquad}
\def\bl{\bigl}
\def\br{\bigr}

\def\IR{\relax{\rm I\kern-.18em R}}

\def\np {  Nucl. Phys. }

%%%%%%%%%%%%%%%%%%%%%%%%%%%%%%%%%%%%%%%%%%%%%%%%%%%%%%%%%%%%%%

\def\r{\rho}
\def\a{\alpha}

\def\b{\beta}

\def\G{\Gamma}
\def\d{\delta}

\def\e{\epsilon}

\def\m{\mu}
\def\n{\nu}
\def\om{\omega}
\def\Om{\Omega}

\def\s{\sigma}

\def\vphi{\varphi}

\def\cL{{\cal L}}

\def\IR{\relax{\rm I\kern-.18em R}}

\def \ha {{1\over 2}}

\def \ov {\over}

%%%%%%%%%%%%%%%%%%%%%%%%%%%%%%%%%%%%%%%%%%%%%%%%%%%%%%%%%%%%%%%%

%references

\lref\BSthree{I. Bars and K. Sfetsos, Mod. Phys. Lett. {\bf A7} (1992) 1091.}

\lref\BShet{I. Bars and K. Sfetsos, Phys. Lett. {\bf 277B} (1992) 269.}

\lref\BSglo{I. Bars and K. Sfetsos, Phys. Rev. {\bf D46} (1992) 4495.}

\lref\BSexa{I. Bars and K. Sfetsos, Phys. Rev. {\bf D46} (1992) 4510.}

\lref\SFET{K. Sfetsos, Nucl. Phys. {\bf B389} (1993) 424.}

\lref\BSslsu{I. Bars and K. Sfetsos, Phys. Lett. {\bf 301B} (1993) 183.}

\lref\BSeaction{I. Bars and K. Sfetsos, Phys. Rev. {\bf D48} (1993) 844.}
% (hep-th/9301047).}

\lref\BN{ I. Bars and D. Nemeschansky, Nucl. Phys. {\bf B348} (1991) 89.}

\lref\BCR{K. Bardakci, M. Crescimanno and E. Rabinovici,
Nucl. Phys. {\bf B344} (1990) 344.}

\lref\WIT{E. Witten, Phys. Rev. {\bf D44} (1991) 314.}

 \lref\IBhet{ I. Bars, Nucl. Phys. {\bf B334} (1990) 125. }

 \lref\IBCS{ I. Bars, ``String Propagation on Black Holes'', USC-91-HEP-B3.\hb
{\it Curved Space-time Strings and Black Holes},
in Proc.
 {\it XX$^{th}$ Int. Conf. on Diff. Geometrical Methods in Physics}, eds. S.
 Catto and A. Rocha, Vol. 2, p. 695, (World Scientific, 1992).}

 \lref\CRE{M. Crescimanno, Mod. Phys. Lett. {\bf A7} (1992) 489.}

\lref\MSW{G. Mandal, A. Sengupta and S. Wadia,
Mod. Phys. Lett. {\bf A6} (1991) 1685.}

 \lref\HOHO{J.B. Horne and G.T. Horowitz, Nucl. Phys. {\bf B368} (1992) 444.}

 \lref\FRA{E. S. Fradkin and V. Ya. Linetsky, Phys. Lett. {\bf 277B}
          (1992) 73.}

 \lref\ISH{N. Ishibashi, M. Li, and A. R. Steif,
         Phys. Rev. Lett. {\bf 67} (1991) 3336.}

 \lref\HOR{P. Horava, Phys. Lett. {\bf 278B} (1992) 101.}

 \lref\RAI{E. Raiten, ``Perturbations of a Stringy Black Hole'',
         Fermilab-Pub 91-338-T.}

 \lref\GER{D. Gershon, Phys. Rev. {\bf D49} (1994) 999.}
%this is: Coset models obtained by twisting WZW models and stringy charged
%black holes in four-dimensions, hepth/9210160.
%this version is wrong: ``Exact Solutions of Four-Dimensional Black Holes in
%String Theory'', TAUP-1937-91, hepth/9202005.}

\lref\GERexa{D. Gershon, ``Semiclassical vs. Exact Solutions of charged Black
Hole in four dimensions and exact $O(D,D)$ duality'', TAUP-2121-93,
hepth/9311122.}

\lref\GERexadual{D. Gershon, ``Exact $O(D,D)$ transformations in WZW models,
TAUP-2129-93, hepth/9312154.}

 \lref \GIN {P. Ginsparg and F. Quevedo,  Nucl. Phys. {\bf B385} (1992) 527. }

 \lref\HOHOS{ J.H. Horne, G.T. Horowitz and A. R. Steif, Phys. Rev. Lett.
 {\bf 68} (1991) 568.}

 \lref\groups{
 M. Crescimanno. Mod. Phys. Lett. {\bf A7} (1992) 489. \hb
 J. B. Horne and G.T. Horowitz, Nucl. Phys. {\bf B368} (1992) 444. \hb
 E. S. Fradkin and V. Ya. Linetsky, Phys. Lett. {\bf 277B} (1992) 73. \hb
 P. Horava, Phys. Lett. {\bf 278B} (1992) 101.\hb
 E. Raiten, ``Perturbations of a Stringy Black Hole'',
         Fermilab-Pub 91-338-T.\hb
 D. Gershon, ``Exact Solutions of Four-Dimensional Black Holes in
         String Theory'', TAUP-1937-91.}

\lref\NAWIT{C. Nappi and E. Witten, Phys. Lett. {\bf 293B} (1992) 309.}

\lref\FRATSE{E. S. Fradkin and A.A. Tseytlin,
Phys. Lett. {\bf 158B} (1985) 316.}

\lref\DASA{ S. Das and B. Sathiapalan, Phys. Rev. Lett. {\bf 56} (1986) 2664.}

\lref\CALLAN{ C.G. Callan, D. Friedan, E.J. Martinec and M. Perry,
Nucl. Phys. {\bf B262} (1985) 593.}

\lref\DB{L. Dixon, J. Lykken and M. Peskin, Nucl. Phys.
{\bf B325} (1989) 325.}

\lref\IB{I. Bars, Nucl. Phys. {\bf B334} (1990) 125.}

\lref\BUSCHER{T. Buscher, Phys. Lett. {\bf 194B} (1087) 59;
Phys. Lett. {\bf 201B} (1988) 466; For reviews see,
A. Giveon, M. Porrati and E. Rabinovici, Phys. Rep. {\bf 244} (1994)
77; E. Alvarez, L. Alvarez-Gaume and Y. Lozano, Nucl. Phys. (Proc. supp.)
{\bf 41} (1995) 1.}

\lref\RV{M. Rocek and E. Verlinde, Nucl. Phys. {\bf B373} (1992) 630.}
\lref\GR{A. Giveon and M. Rocek, Nucl. Phys. {\bf B380} (1992) 128. }

\lref\nadual{X. C. de la Ossa and F. Quevedo,
Nucl. Phys. {\bf B403} (1993) 377.}

\lref\duearl{B.E. Fridling and A. Jevicki,
Phys. lett. {\bf B134} (1984) 70.\hb
E.S. Fradkin and A.A. Tseytlin, Ann. Phys. {\bf 162} (1985) 31.}

\lref\ABEN{I. Antoniadis, C. Bachas, J. Ellis and D.V. Nanopoulos,
Phys. Lett. {\bf B211} (1988) 393.}

\lref\Kalofour{N. Kaloper, Phys. Rev. {\bf D48} (1993) 4658.}
%Exact primordial black strings in four-dimensions, hepth/9303059.

\lref\GPS{S.B. Giddings, J. Polchinski and A. Strominger,
Phys. Rev. {\bf D48} (1993) 5784.}
%Four-Dimensional black holes in string theory, hepth/9305083.

\lref\SENrev{A. Sen, ``Black Holes and Solitons in String Theory'',
TIFR-TH-92-57.}
% October 1992.}

\lref\TSEd{A.A. Tseytlin, Mod. Phys. Lett. {\bf A6} (1991) 1721.}

\lref\TSESC{A. S. Schwarz and A.A. Tseytlin, ``Dilaton shift under duality
and torsion of elliptic complex'', IMPERIAL/TP/92-93/01. }

\lref\Dualone{K. Meissner and G. Veneziano,
Phys. Lett. {\bf B267} (1991) 33;
Mod. Phys. Lett. {\bf A6} (1991) 3397. \hb
M. Gasperini and G. Veneziano, Phys. Lett. {\bf 277B} (1992) 256. \hb
M. Gasperini, J. Maharana and G. Veneziano, Phys. Lett. {\bf 296B} (1992) 51.}

\lref\rovr{K. Kikkawa and M. Yamasaki, Phys. Lett. {\bf B149} (1984) 357.\hb
N. Sakai and I. Senda, Prog. theor. Phys. {\bf 75} (1986) 692.}

\lref\narain{K.S. Narain, Phys. Lett. {\bf B169} (1986) 369.\hb
K.S. Narain, M.H. Sarmadi and C. Vafa, Nucl. Phys. {\bf B288} (1987) 551.}

\lref\GV{P. Ginsparg and C. Vafa, Nucl. Phys. {\bf B289} (1987) 414.}
\lref\nssw{V. Nair, A. Shapere, A. Strominger and F. Wilczek,
Nucl. Phys. {\bf B287} (1987) 402.}

\lref\vafa{C. Vafa, ``Strings and Singularities'', HUTP-93/A028.}

\lref\Dualtwo{A. Sen,
Phys. Lett. {\bf B271} (1991) 295;\ ibid. {\bf B274} (1992) 34;
Phys. Rev. Lett. {\bf 69} (1992) 1006. \hb
S. Hassan and A. Sen, Nucl. Phys. {\bf B375} (1992) 103. \hb
J. Maharana and J. H. Schwarz, Nucl. Phys. {\bf B390} (1993) 3.\hb
A. Kumar, Phys. Lett. {\bf B293} (1992) 49.}

\lref\dualmargi{S.F. Hassan and A. Sen, Nucl. Phys. {\bf B405} (1993) 143;
M. Henningson and C. Nappi, Phys. Rev. {\bf D48} (1993) 861;
E. Kiritsis, Nucl. Phys. {\bf B405} (1993) 109.}

\lref\bv{R. Brandenberger and C. Vafa, Nucl. Phys. {\bf B316} (1989) 301.}
\lref\gsvy{B.R. Greene, A. Shapere, C. Vafa and S.T. Yau,
Nucl. Phys. {\bf B337} (1990) 1.}
\lref\tv{A.A. Tseytlin and C. Vafa,  Nucl. Phys. {\bf B372} (1992) 443.}

\lref\grvm{A. Shapere and F. Wilczek, Nucl. Phys. {\bf B320} (1989) 609.\hb
A. Giveon, E. Rabinovici and G. Veneziano,
Nucl. Phys. {\bf B322} (1989) 167.\hb
A. Giveon, N. Malkin and E. Rabinovici, Phys. Lett. {\bf B238} (1990) 57.}

\lref\GRV{M. Gasperini, R. Ricci and G. Veneziano,
Phys. Lett. {\bf B319} (1993) 438.}
%\hepth 9308112

\lref\KIRd{E. Kiritsis, Nucl. Phys. {\bf B405} (1993) 109.}
%``Exact Duality Symmetries in CFT and String Theory'',
%LPTENS-92-29, CERN-TH-6797-93.}
% February 1993.}

%\lref\dualrev{For reviews see,
%A. Giveon, M. Porrati and E. Rabinovici, Phys. Rep. {\bf 244} (1994)
%77; E. Alvarez, L. Alvarez-Gaume and Y. Lozano, Nucl. Phys. (Proc. supp.)
%{\bf 41} (1995) 1.}

%``Target Space Duality
%in String Theory'', RI-1-94, hepth/9401139.}

\lref\slt{A. Giveon, Mod. Phys. Lett. {\bf A6} (1991) 2843.}

\lref\GIPA{A. Giveon and A. Pasquinucci, ``On cosmological string backgrounds
with toroidal isometries'', IASSNS-HEP-92/55, August 1992.}

\lref\KASU{Y. Kazama and H. Suzuki, Nucl. Phys. {\bf B234} (1989) 232. \hb
Y. Kazama and H. Suzuki Phys. Lett. {\bf 216B} (1989) 112.}

\lref\WITanom{E. Witten, Comm. Math. Phys. {\bf 144} (1992) 189.}

\lref\WITnm{E. Witten, Nucl. Phys. {\bf B371} (1992) 191.}

\lref\IBhetero{I. Bars, Phys. Lett. {\bf 293B} (1992) 315.}

\lref\IBerice{I. Bars, {\it Superstrings on Curved Space-times}, Lecture
delivered at the Int. workshop on {\it String Quantum Gravity and Physics
at the Planck Scale}, Erice, Italy, June 1992.}

\lref\DVV{R. Dijkgraaf, E. Verlinde and H. Verlinde, Nucl. Phys. {\bf B371}
(1992) 269.}

\lref\TSEY{A.A. Tseytlin, Phys. Lett. {\bf 268B} (1991) 175.}

\lref\JJP{I. Jack, D. R. T. Jones and J. Panvel,
          Nucl. Phys. {\bf B393} (1993) 95.}

\lref\BST { I. Bars, K. Sfetsos and A.A. Tseytlin, unpublished. }

\lref\TSEYT{ A.A. Tseytlin, Nucl. Phys. {\bf B399} (1993) 601.}

\lref\TSEYTt{A.A. Tseytlin, Nucl. Phys. {\bf B411} (1993) 509.}
%``Conformal Sigma Models corresponding to Gauged
%WZW Models'', CERN-TH.6804/93, hepth/9302083.}

 \lref\SHIF { M. A. Shifman, Nucl. Phys. {\bf B352} (1991) 87.}
\lref\SHIFM { H. Leutwyler and M. A. Shifman, Int. J. Mod. Phys. {\bf
A7} (1992) 795. }

%\lref\POLWIG { A. M. Polyakov and P. B. Wiegman, Phys.
%Lett. {\bf 131B} (1984) 121.  }
\lref\POLWIG { A. M. Polyakov and P. B. Wiegman, Phys.
Lett. {\bf 141B} (1984) 223.  }

\lref\BCR{K. Bardakci, M. Crescimanno
and E. Rabinovici, Nucl. Phys. {\bf B344} (1990) 344. }

\lref\Wwzw{E. Witten, Commun. Math. Phys. {\bf 92} (1984) 455.}

\lref\GKO{P. Goddard, A. Kent and D. Olive, Phys. Lett. {\bf 152B} (1985) 88.}

\lref\Toda{A. N. Leznov and M. V. Saveliev, Lett. Math. Phys. {\bf 3} (1979)
489. \hb A. N. Leznov and M. V. Saveliev, Comm. Math. Phys. {\bf 74}
(1980) 111.}

\lref\GToda{J. Balog, L. Feh\'er, L. O'Raifeartaigh, P. Forg\'acs and A. Wipf,
Ann. Phys. (New York) {\bf 203} (1990) 76; Phys. Lett. {\bf 244B}
(1990) 435.}

\lref\GWZW{ E. Witten, \np {\bf B223} (1983) 422. \hb
K. Bardakci, E. Rabinovici and B. S\"aring, Nucl. Phys. {\bf B299}
(1988) 157. \hb K. Gawedzki and A. Kupiainen, Phys. Lett. {\bf 215B}
(1988) 119.; Nucl. Phys. {\bf B320} (1989) 625. }

\lref\SCH{ D. Karabali, Q-Han Park, H. J. Schnitzer and Z. Yang,
                   Phys. Lett. {\bf B216} (1989) 307. \hb D. Karabali
and H. J. Schnitzer, Nucl. Phys. {\bf B329} (1990) 649. }

 \lref\KIR{E. Kiritsis, Mod. Phys. Lett. {\bf A6} (1991) 2871. }

\lref\BIR{N. D. Birrell and P. C. W. Davies,
{\it Quantum Fields in Curved Space}, Cambridge University Press.}

\lref\WYB{B. G. Wybourn, {\it Classical Groups for Physicists }
(John Wiley \& sons, 1974).}

\lref\Brinkman{H.W. Brinkmann, Math. Ann. {\bf 94} (1925) 119.}

\lref\SANTA{R. Guven, Phys. Lett. {\bf 191B} (1987) 275.\hb
D. Amati and C. Klimcik, Phys. Lett. {\bf 219B} (1989) 443.\hb
G.T. Horowitz and A. R. Steif, Phys. Rev. Lett. {\bf 64} (1990) 260;
Phys. Rev. {\bf D42} (1990) 1950.\hb
R.E. Rudd, Nucl. Phys. {\bf B352} (1991) 489.\hb
C. Duval, G.W. Gibbons and P.A. Horvathy, Phys. Rev. {\bf D43} (1991) 3907.\hb
C. Duval, Z. Horvath and P.A. Horvathy, Phys. Lett. {\bf B313} (1993) 10.\hb
E.A. Bergshoeff, R. Kallosh and T. Ortin, Phys. Rev. {\bf D47} (1993) 5444.}
\lref\SANT{J. H. Horne, G.T. Horowitz and A. R. Steif,
Phys. Rev. Lett. {\bf 68} (1991) 568.}

\lref\tsecov{A.A. Tseytlin, Nucl. Phys. {\bf B390} (1993) 153;
Phys. Rev. {\bf D47} (1993) 3421.}

\lref\garriga{J. Garriga and E. Vardaguer, Phys. Rev. {\bf D43} (1991) 391.}

\lref\PRE{J. Prescill, P. Schwarz, A. Shapere, S. Trivedi and F. Wilczek,
Mod. Phys Lett. {\bf A6} (1991) 2353.\hb
C. Holzhey and F. Wilczek, Nucl. Phys. {\bf B380} (1992) 447.}

\lref\HAWK{J. B. Hartle and S. W. Hawking Phys. Rev. {\bf D13} (1976) 2188.\hb
S. W. Hawking, Phys. Rev. {\bf D18} (1978) 1747.}

\lref\HAWKI{S. W. Hawking, Comm. Math. Phys. {\bf 43} (1975) 199.}

\lref\HAWKII{S. W. Hawking, Phys. Rev. {\bf D14} (1976) 2460.}

\lref\euclidean{S. Elitzur, A. Forge and E. Rabinovici,
Nucl. Phys. {\bf B359} (1991) 581. }

\lref\ITZ{C. Itzykson and J. Zuber, {\it Quantum Field Theory},
McGraw Hill (1980). }

\lref\kacrev{P. Goddard and D. Olive, Journal of Mod. Phys. {\bf A} Vol. 1,
No. 2 (1986) 303.}

\lref\BBS{F.A. Bais, P. Bouwknegt, K.S. Schoutens and M. Surridge,
Nucl. Phys. {\bf B304} (1988) 348.}

\lref\nonl{A. Polyakov, {\it Fields, Strings and Critical Phenomena}, Proc. of
Les Houses 1988, eds. E. Brezin and J. Zinn-Justin North-Holland, 1990.\hb
Al. B. Zamolodchikov, preprint ITEP 87-89. \hb
K. Schoutens, A. Sevrin and P. van Nieuwenhuizen, Proc. of the Stony Brook
Conference {\it Strings and Symmetries 1991}, World Scientific,
Singapore, 1992. \hb
J. de Boer and J. Goeree, ``The Effective Action of $W_3$ Gravity to all
\hb orders'', THU-92/33.}

\lref\HOrev{G.T. Horowitz, {\it The Dark Side of String Theory:
Black Holes and Black Strings}, Proc. of the 1992 Trieste Spring School on
String Theory and Quantum Gravity.}

\lref\HSrev{J. Harvey and A. Strominger, {\it Quantum Aspects of Black
Holes}, Proc. of the 1992 Trieste Spring School on
String Theory and Quantum Gravity.}

\lref\GM{G. Gibbons, Nucl. Phys. {\bf B207} (1982) 337.\hb
G. Gibbons and K. Maeda, Nucl. Phys. {\bf B298} (1988) 741.}

\lref\GID{S. B. Giddings, Phys. Rev. {\bf D46} (1992) 1347.}

\lref\PRErev{J. Preskill, {\it Do Black Holes Destroy Information?},
Proc. of the International Symposium on Black Holes, Membranes, Wormholes,
and Superstrings, The Woodlands, Texas, 16-18 January, 1992.}

\lref\tye{S-W. Chung and S. H. H. Tye, Phys. Rev. {\bf D47} (1993) 4546.}

\lref\eguchi{T. Eguchi, Mod. Phys. Lett. {\bf A7} (1992) 85.}

\lref\blau{M. Blau and G. Thompson, Nucl. Phys. {\bf B408} (1993) 345.}

\lref\HSBW{P. S. Howe and G. Sierra, Phys. Lett. {\bf 144B} (1984) 451.\hb
J. Bagger and E. Witten, Nucl. Phys. {\bf B222} (1983) 1.}

\lref\GSW{M. B. Green, J. H. Schwarz and E. Witten, {\it Superstring Theory},
Cambridge Univ. Press, Vols. 1 and 2, London and New York (1987).}

\lref\KAKU{M. Kaku, {\it Introduction to Superstrings}, Springer-Verlag, Berlin
and New York (1991).}

\lref\LSW{W. Lerche, A. N. Schellekens and N. P. Warner, {\it Lattices and
Strings }, Physics Reports {\bf 177}, Nos. 1 \& 2 (1989) 1, North-Holland,
Amsterdam.}

\lref\confrev{P. Ginsparg and J. L. Gardy in {\it Fields, Strings, and
Critical Phenomena}, 1988 Les Houches School, E. Brezin and J. Zinn-Justin,
eds, Elsevier Science Publ., Amsterdam (1989). \hb
J. Bagger, {\it Basic Conformal Field Theory},
Lectures given at 1988 Banff Summer Inst. on Particle and Fields,
Banff, Canada, Aug. 14-27, 1988, HUTP-89/A006, January 1989. }

\lref\CHAN{S. Chandrasekhar, {\it The Mathematical Theory of Black Holes},
Oxford University Press, 1983.}

\lref\KOULU{C. Kounnas and D. L\"ust, Phys. Lett. {\bf 289B} (1992) 56.}

\lref\PERRY{M. J. Perry and E. Teo, Phys. Rev. Lett. {\bf 70} (1993) 2669.\hb
P. Yi, Phys. Rev. {\bf D48} (1993) 2777.}
%``Nonsingular 2d Black Holes and Classical String Backgrounds'',

\lref\GiKi{A. Giveon and E. Kiritsis, Nucl. Phys. {\bf B411} (1994) 487.}
%``Axial Vector Duality as a gauge
%Symmetry and Topology Change in String Theory'', CERN-TH-6816-93,
%hepth/9303016, to appear in Nucl. Phys. {\bf B} (1994).}
% February 1993.}

\lref\kar{S.K. Kar and A. Kumar, Phys. Lett. {\bf 291B} (1992) 246.}

\lref\NW{C. Nappi and E. Witten, Phys. Rev. Lett. {\bf 71} (1993) 3751.}
%``A WZW model based on a non-semi-simple group'',
%IASSNS-HEP-93/61, hepth/9310112.}

\lref\HK{M. B. Halpern and E. Kiritsis,
Mod. Phys. Lett. {\bf A4} (1989) 1373.}

\lref\MOR{A.Yu. Morozov, A.M. Perelomov, A.A. Rosly, M.A. Shifman and
A.V. Turbiner, Int. J. Mod. Phys. {\bf A5} (1990) 803.}

\lref\KK{E. Kiritsis and C. Kounnas, Phys. Lett. {\bf B320} (1994) 264.}
%``String propagation in Gravitational Wave Backgrounds'',
%CERN-TH.7059/93, hepth/9310202.}

\lref\KKdyn{E. Kiritsis and C. Kounnas, Phys. Lett. {\bf B331} (1994) 51.}

\lref\Tsdyn{A.A. Tseytlin, Phys. Lett. {\bf B334} (1994) 315.}

\lref\KST{K. Sfetsos and A.A. Tseytlin, Phys. Rev. {\bf D49} (1994) 2933.}
%``Antisymmetric tensor coupling and conformal
%invariance in sigma models corresponding to gauged WZNW theories'',
%THU-93/25, CERN-TH.6969/93, hepth/9310159,
%to appear in Phys. Rev. {\bf D} (1994).}

\lref\KSTh{K. Sfetsos and A.A. Tseytlin, Nucl. Phys. {\bf B415} (1994) 116.}
%``Chiral gauged WZNW models and heterotic string
%backgrounds'', CERN-TH.6962/93, USC-93/HEP-S2, hepth/9308018,
%to appear in Nucl. Phys. {\bf B} (1994).}

\lref\KP{S.P. Khastgir and A. Kumar, ``Singular limits and string solutions'',
IP/BBSR/93-72, hepth/9311048.}

\lref\etc{K. Sfetsos, Phys. Lett. {\bf B324} (1994) 335.}
%``Gauging a non-semi-simple WZW model'',
%THU-93/30, hepth/9311010 (revised), to appear in Phys. Lett. {\bf B} (1994).}

\lref\KTone{C. Klimcik and A.A. Tseytlin, Phys. Lett. {\bf B323} (1994) 305.}
%``Duality invariant class of
%exact string backgrounds'', CERN TH.7069, hepth/9311012.}

\lref\KTtwo{C. Klimcik and A.A. Tseytlin, ``Exact four dimensional string
solutions and Toda-like sigma models from `null-gauged' WZNW theories,
Imperial/TP/93-94/17, PRA-HEP 94/1, hepth/9402120.}

\lref\saletan{E.J. Saletan, J. Math. Phys. {\bf 2} (1961) 1.}
\lref\jao{D. Cangemi and R. Jackiw, Phys. Rev. Lett. {\bf 69} (1992) 233.}
\lref\jat{D. Cangemi and R. Jackiw, Ann. Phys. (NY) {\bf 225} (1993) 229.}

\lref\ORS{ D. I. Olive, E. Rabinovici and A. Schwimmer, Phys. Lett. {\bf B321}
(1994) 361.}
%``A class of String
%Backgrounds as a Semiclassical Limit of WZW Models'', SWA/93-94/15,
%hepth/93011081.}

\lref\edc{K. Sfetsos, ``Exact String Backgrounds from WZW models based on
Non-semi-simple groups'', THU-93/31, hepth/9311093, to appear in
Int. J. Mod. Phys. {\bf A} (1994).}

\lref\sfedual{K. Sfetsos, Phys. Rev. {\bf D50} (1994) 2784.}
%``Gauged WZW models and Non-abelian Duality'',\hb
%THU-94/01, hepth/9402031.}

\lref\grnonab{A. Giveon and M. Rocek, Nucl. Phys. {\bf B421} (1994) 173.}
% ``On non-abelian duality'',
%ITP-SB-93-44, RI-152-93, hepth/9308154.}

\lref\wigner{E. ${\rm In\ddot on\ddot u}$ and E.P. Wigner,
Proc. Natl. Acad. Sci. U. S. {\bf 39} (1953) 510.}

\lref\HY{ J. Yamron and M.B. Halpern, Nucl. Phys. {\bf B351} (1991) 333.}
\lref\HB{K. Bardakci and M.B. Halpern, Phys. Rev. {\bf D3} (1971) 2493.}
\lref\MBH{M.B. Halpern, Phys. Rev {\bf D4} (1971) 2398.}
\lref\balog{J. Balog, L. O'Raifeartaigh, P. Forgacs and A. Wipf,
Nucl. Phys. {\bf B325} (1989) 225.}
\lref\kacm{V. G. Kac, Funct. Appl. {\bf 1} (1967) 328.\hb
R.V. Moody, Bull. Am. Math. Soc. {\bf 73} (1967) 217.}

\lref\TSEma{A.A. Tseytlin, Nucl. Phys. {\bf B418} (1994) 173.}
%``On a Universal class of WZW-type
%conformal models'', CERN-TH.7068/93, hepth/9311062.}

\lref\BCHma{J. de Boer, K. Clubock and M.B. Halpern, ``Linearized form
of the Generic Affine-Virasoro Action'', UCB-PTH-93/34, hepth/9312094.}

\lref\AABL{E. Alvarez, L.Alvarez-Gaume, J.L.F. Barbon and Y. Lozano,
Nucl. Phys. {\bf B415} (1994) 71.}
%``Some global aspects of Duality in String Theory'', CERN-TH.6991/93,
%FTUAM.93/28, hepth/9309039.}

\lref\noumo{N. Mohammedi,  Phys. Lett. {\bf B325} (1994) 371.}
%``On Bosonic and Superconformal Current Algebra
%for Non-semi-simple Groups'', BONN-HE-93-51, hepth/9312182.}

\lref\sfetse{K. Sfetsos and A.A. Tseytlin, unpublished (February 1994).}
\lref\figsonia{J.M. Figueroa-O'Farrill and S. Stanciu, Phys. Lett. {\bf B327}
(\1994) 40.}
%``Nonsemisimple Sugawara constructions'', QMW-PH-94-2, hepth/9402035.}

\lref\ALLleadord{E. Witten, Phys. Rev. {\bf D44} (1991) 314.\hb
J.B. Horne and G.T. Horowitz, Nucl. Phys. {\bf B368} (1992) 444.\hb
M. Crescimanno, Mod. Phys. Lett. {\bf A7} (1992) 489.\hb
I. Bars and K. Sfetsos, Mod. Phys. Lett. {\bf A7} (1992) 1091.\hb
E.S. Fradkin and V.Ya. Linetsky, Phys. Lett. {\bf 277B} (1992) 73. \hb
I. Bars and K. Sfetsos, Phys. Lett. {\bf 277B} (1992) 269.\hb
P. Horava, Phys. Lett. {\bf 278B} (1992) 101.\hb
E. Raiten, ``Perturbations of a Stringy Black Hole'', Fermilab-Pub 91-338-T.\hb
P. Ginsparg and F. Quevedo,  Nucl. Phys. {\bf B385} (1992) 527.\hb
more}

\lref\ALLexact{
R. Dijkgraaf, E. Verlinde and H. Verlinde, Nucl. Phys. {\bf B371} (1992) 269.
\hb
I. Bars and K. Sfetsos, Phys. Rev. {\bf D46} (1992) 4510.;
Phys. Lett. {\bf 301B} (1993) 183.\hb
K. Sfetsos, Nucl. Phys. {\bf B389} (1993) 424.\hb
Gerhon}
%%%%%%%%%%%%%%%%%%%%%%%%%%%%%%%%%%%%%%%%%%%%%%%%

\lref\DHooft{T. Dray and G. 't Hooft, Nucl. Phys. {\bf B253} (1985)
173.}
\lref\MyPe{ R.C. Myers and M.J. Perry, Ann. of Phys. {\bf 172} (1986)
304.}
\lref\GiMa{G. Gibbons and K. Maeda, Nucl. Phys. {\bf B298} (1988) 741.}
\lref\KLOPV{R. Kallosh, A. Linde, T. Ort\'\i n, A. Peet and A. Van Proeyen,
Phys. Rev. {\bf D46} (1992) 5278.}

\lref\lousan{ C.O. Loust\'o and N. S\'anchez, Int. J. Mod. Phys. {\bf A5}
(1990) 915; Nucl. Phys. {\bf B355} (1991) 231.\hb
V. Ferrari and P. Pendenza, Gen. Rel. Grav. {\bf 22} (1990)
1105.\hb
H. Balasin and H. Nachbagauer, Class. Quantum Grav. {\bf 12} (1995) 707. \hb
%``The Ultrarelativistic Kerr-Geometry and
%its Energy-Momentum Tensor'', TUW-94-09, gr-qc/9405053.
K. Hayashi and T. Samura, Phys. Rev. {\bf D50} (1994) 3666. \hb
S. Das and P. Majumdar, Phys. Lett. {\bf B348} (1995) 349.}
%``Shock wave mixing in Einstein and Dilaton Gravity'',
%IMSc-94/46, hepth/9411129.}

%\lref\lousanII{ C.O. Loust\'o and N. S\'anchez, Nucl. Phys. {\bf B383} (1992)
%377.}

\lref\lousanII{ C.O. Loust\'o and N. S\'anchez, ``Scattering processes at the
Planck scale'', UAB-FT-353, gr-qc/9410041 and references therein.}

\lref\DHooftII{ T. Dray and G. 't Hooft, Commun. Math. Phys. {\bf 99}
(1985) 613.}
\lref\DHooftIII{T. Dray and G. 't Hooft,
Class. Quant. Grav. {\bf 3} (1986) 825.}

\lref\DHooftII{ T. Dray and G. 't Hooft, Commun. Math. Phys. {\bf 99}
(1985) 613; Class. Quant. Grav. {\bf 3} (1986) 825.}

\lref\VV{H. Verlinde and E. Verlinde, Nucl. Phys. {\bf B371} (1992)
246.}

\lref\aisexl{P.C. Aichelburg and R.U. Sexl, Gen. Rel. Grav. {\bf 2}
(1971) 303.}
\lref\cagan{C.G. Callan and Z. Gan, Nucl. Phys. {\bf B272} (1986) 647.}
\lref\hooft{ G. 't Hooft, Nucl. Phys. {\bf B335} (1990) 138.}
\lref\hoo{G. 't Hooft, Phys. Lett. {\bf B198} (1987) 61;
Nucl. Phys. {\bf B304} (1988) 867.}

\lref\tasepr{ I.S. Gradshteyn and I.M. Ryznik, {\it Tables of integrals,
Series and Products}, Academic, New York, (1980). }
\lref\HOTA{ M. Hotta and M. Tanaka, Class. Quantum Grav. {\bf 10} (1993) 307.}
\lref\otth{ V. Ferrari and P. Pendenza, Gen. Rel. Grav. {\bf 22} (1990)
1105.\hb
H. Balasin and H. Nachbagauer, Class. Quantum Grav. {\bf 12} (1995) 707. \hb
%``The Ultrarelativistic Kerr-Geometry and
%its Energy-Momentum Tensor'', gr-qc/9405053.
K. Hayashi and T. Samura, Phys. Rev. {\bf D50} (1994) 3666.}
%``Gravitational shock waves for Schwarzschild and Kerr black holes'',
%KOBE-FHD-93-03A, gr-gc/9404027.}
%``Planckian scattering of massive particles and gravitational
%shock waves'', KOBE-FHD-93-05A, hep-th/9405013.}

\lref\KSHM{ D. Kramer, H. Stephani, E. Herlt and M. MacCallum,
{\it Exact Solutions of Einstein's Field Equations}, Cambridge (1980).}
%\lref\lou{ C.O. Lousto, ``The emergence of an effective two-dimensional
%Quantum Description from the study of Critical Phenomena in Black Holes'',
%gr-gc/9405048.}

\lref\lathos{ C.O. Loust\'o and N. S\'anchez, Phys. Lett. {\bf B220} (1989)
55.}

\lref\BAIT{ M. Bander and C. Itzykson, Rev. of Mod. Phys. {\bf 38} (1966) 330;
Rev. of Mod. Phys. {\bf 38} (1966) 346.}

\lref\rpen{R. Penrose, in General Relativity: papers in honour of J.L. Synge,
ed. L. O'Raifeartaigh (Clarendon, Oxford, 1972) 101.}
\lref\AGV{L.N. Lipatov, Nucl. Phys. {\bf B365} (1991) 614.\hb
R. Jackiw, D. Kabat and M. Ortiz,
Phys. Lett. {\bf B277} (1992) 148.\hb
D. Kabat and M. Ortiz, Nucl. Phys. {\bf B388} (1992) 570.\hb
D. Amati, M. Ciafaloni and G. Veneziano,
Nucl. Phys. {\bf B403} (1993) 707.}
%M. Fabbrichesi, R. Pettorino, G. Veneziano and G.A. Vilkovisky,
%Nucl. Phys. {\bf B419} (1994) 147.}

%F.M. Paiva, M.J. Reboukas and M.A.H. MacCallum,
%Class. Quant. Grav. {\bf 10} (1993) 1165;

%%%%%%%%%%%%%%%%%%%%%%%%%%%%%%%%%%%%%%%%%%%%%%%%%%%%%%%%%%%%%%%%%%%%%%
\lref\KPen{ K.A. Khan and R. Penrose, Nature (London) {\bf 229} (1971) 185.}
\lref\DOVE{M. Dorca and E. Verdaguer, Nucl. Phys. {\bf B403} (1993) 770.}

\lref\xanth{S. Cahndrasekhar and B. Xanthopoulos }

\lref\stplane{K. Sfetsos and A.A. Tseytlin, Nucl. Phys. {\bf B427} (1994) 245.}

\lref\shock{ K. Sfetsos, Nucl. Phys. {\bf B436} (1995) 721.}
%``On gravitational shock waves on curved spacetimes'',
%THU-94/13, hepth/9408169.}

\lref\gardiner{C.W. Gardiner, {\it Handbook of stochastic methods for Physics,
Chemistry and the Natural sciences}, Berlin, Springer, 1983.}

\lref\CHSW{ S. Chaudhuri and J.A. Schwarz, Phys. Lett. {\bf B219} (1989) 291.}

\lref\DJHOT{ D.H. Hartley, M. \"Onder and R.W. Tucker,
Class. Quantum Grav. {\bf 6} (1989) 1301\hb
T. Dray and P. Joshi, Class. Quantum Grav. {\bf 7} (1990) 41.}

\lref\CHS{ C.G. Callan, J.A. Harvey and A. Strominger, Nucl. Phys. {\bf B359}
(1991) 611.}

\lref\schtalk{ J.H. Schwarz,
``Evidence for Non-perturbative String Symmetries'',
CALT-68-1965, hepth/9411178.}

\lref\bakasII{I. Bakas, Phys. Lett. {\bf B343} (1995) 103.}
%``Space Time Interpretation of S-Duality and
%Supersymmetry Violations of T-Duality'', CERN-TH.7473/94, hepth/9410104.}
%%%%%%%%%%%%%%%%%%%%%%%%
\lref\basfep{I. Bakas and K. Sfetsos, Phys. Lett. {\bf B349} (1995) 448.}
\lref\AALcan{E. Alvarez, L. Alvarez--Gaume and Y. Lozano, Phys. Lett.
{\bf B336} (1994) 183.}
\lref\basusy{ I. Bakas, Phys. Lett. {\bf B343} (1995) 103.}
\lref\BKO{ E. Bergshoeff, R. Kallosh and T. Ortin, Phys. Rev. {\bf D51} (1995)
3009.}
%{\em ``Duality Versus
%Supersymmetry and Compactification"}, Groningen preprint UG--8--94,
%hep--th/9410230, October 1994.
\lref\KAFK{C. Kounnas, Phys. Lett. {\bf B321} (1994) 26;
I. Antoniadis, S. Ferrara and C. Kounnas, Nucl. Phys. {\bf B421} (1994) 343.}

\lref\ALFR{ L. Alvarez--Gaume and D. Freedman, Commun. Math. Phys.
{\bf 80} (1981) 443.}
\lref\GHR{S. Gates, C. Hull and M. Rocek, Nucl. Phys. {\bf B248} (1984)
157.}
\lref\HOPAPA{P. Howe and G. Papadopoulos, Nucl. Phys. {\bf B289} (1987) 264;
Class. Quant. Grav. {\bf 5} (1988) 1647.}

\lref\zumino{B. Zumino, Phys Lett. {\bf B87} (1979) 203.}

\lref\PNBW{P. van Nieuwenhuizen
and B. de Wit, Nucl. Phys. {\bf B312} (1989) 58.}
\lref\CHS{ C. Callan, J. Harvey and A. Strominger, Nucl. Phys.
{\bf B359} (1991) 611.}

\lref\hassand{S.F. Hassan ``T--Duality and Non--local Supersymmetries'',
CERN-TH/95-98,hep-th/9504148.}

\lref\AAB{E. Alvarez, L. Alvarez-Gaume and I. Bakas, ``T--Duality and
Space-time Supersymmetry", CERN-TH-95-201, hep-th/9507112;
``Supersymmetry and Dualities'', Contribution to the Trieste Conference
on {\it Mirror symmetry and S--Duality}, Trieste June 5-9, 1995,
CERN-TH-95-258, hep-th/9510028.}

\lref\IKR{I. Ivanov, B. Kim and M. Rocek, Phys. Lett. {\bf B343} (1995).}
\lref\BOVA{G. Bonneau and G. Valent, Class. Quant. Grav. {\bf 11} (1994) 1133.}
\lref\GIRU{ G. Gibbons and P. Ruback, Commun. Math. Phys. {\bf 115}
(1988) 267.}

\lref\BOFI{ C. Boyer and J. Finley, J. Math. Phys. {\bf 23} (1982) 1126;
J. Gegenberg and A. Das, Gen. Rel. Grav. {\bf 16} (1984) 817.}
\lref\HAGITO{ S. Hawking, Phys. Lett. {\bf A60} (1977) 81;
G. Gibbons and S. Hawking, Commun. Math. Phys. {\bf 66}
(1979) 291;
K. Tod and R. Ward, Proc. R. Soc. Lond. {\bf A368} (1979) 411.}

\lref\zachos{ T.L. Curtright and C.K. Zachos, Phys. Rev. Lett. {\bf 53}
 (1984) 1799;
E. Braaten, T.L. Curtright and C.K. Zachos, Nucl. Phys. {\bf B260} (1985) 630.}

\lref\zacdual{T.L. Curtright and
C.K. Zachos, Phys. Rev. {\bf D49} (1994) 5408;
Y. Lozano, Phys. Lett. {\bf B355} (1995) 165.}

\lref\zasusydu{ T.L. Curtright and
C.K. Zachos, Phys. Rev. {\bf D52} (1995) R573.}% hepth/9502126

\lref\FRTO{D.Z. Freedman and P.K. Townsend, Nucl. Phys. {\bf B177} (1981) 282.}
\lref\HS{P. S. Howe and G. Sierra, Phys. Lett. {\bf 144B} (1984) 451.}

\lref\hassanfi{S. Hassan, ``O(D,D;R) Deformations of Complex Structures
and Extended World Sheet Supersymmetry'', TIFR--TH--94--26,
hep--th/9408060.}

\lref\spindel{Ph. Spindel, A. Sevrin, W. Troost and A.Van Proeyen, Nucl.
Phys. {\bf B308} (1988) 662; Nucl. Phys. {\bf B311} (1988/89) 465.}

\lref\STP{ A. Sevrin, W. Troost and A. van Proeyen, Phys. Lett.
{\bf B208} (1988) 447.}

\lref\RSS{ M. Rocek, K. Schoutens and A. Sevrin, Phys. Lett.
{\bf B265} (1991) 303; E. Kiritsis, C. Kounnas and D. Lust,
Int. J. Mod. Phys. {\bf A9} (1994) 1361.}

%\lref\KKL{E. Kiritsis, C. Kounnas and D. Lust,
%Int. J. Mod. Phys. {\bf A9} (1994) 1361.}

\lref\WARNPN{ P. van Nieuwenhuizen and N.P. Warner, Comm. Math. Phys. {\bf 93}
(1984) 277.}

\lref\kilcom{ P. Candelas, G. Horowitz, A. Strominger and E. Witten,
Nucl. Phys.
{\bf B258} (1985) 46; A. Strominger, Nucl. Phys. {\bf B274} (1986) 253.}

\lref\canerl{A. Giveon, E. Rabinovici and G. Veneziano, Nucl. Phys.
{\bf B322} (1989) 167;
K. Meissner and G. Veneziano, Phys. Lett. {\bf B267} (1991) 33.}

\lref\hucom{ C. Hull, Mod. Phys. Lett. {\bf A5} (1990) 1793;
C. Hull and B. Spence, Nucl. Phys. {\bf B345} (1990) 493.}

\lref\basfewopr{ I. Bakas and K. Sfetsos, work in progress.}

\lref\EGHgra{T. Eguchi, P. Gilkey and A. Hanson, Phys. Rep.
{\bf 66} (1980) 213.}

\lref\sferesto{K. Sfetsos,
``Duality and Restoration of Manifest Supersymmetry'', THU-95/20,
hepth/9510034.}

%\lref\lozdu{ Y. Lozano, Phys. Lett. {\bf B355} (1995) 165.}

%%%%%%%%%%%%%%%%%%%%%%%%%%%%%%%%%%%%%%%%%%%%%%%%%%%%%%%%%%%%%%%%
% begin paper

\hfill {THU-95/21}
\vskip -.15 true cm
\rightline{October 1995}
\vskip -.15 true cm
\rightline {hep-th/9510103}

\bs\bs\bs

\centerline { {\bf ON T--DUALITY AND SUPERSYMMETRY}\footnote{$^\dagger$}
{To appear in the proceedings of the Conference on Gauge Theories,
Applied Supersymmetry \vskip -.15 true cm
and Quantum Gravity, Leuven, 10-14 July 1995.}}

\vskip 1.5 true cm

\centerline  {  {\bf Konstadinos Sfetsos}\footnote{$^*$}
{e-mail address: sfetsos@fys.ruu.nl }}

\bigskip

\centerline {Institute for Theoretical Physics }
\centerline {Utrecht University}
\centerline {Princetonplein 5, TA 3508}
\centerline{ The Netherlands }

%%%%%%%%%%%%%%%%%%%%%%%%%%%%%%%%%%%%%%%%%%%%%%%%%%%%%%%%%%%%%%%%%%%%

\vskip 1.50 true cm

\centerline{ABSTRACT}

\bs

The interplay between T--duality and supersymmetry in string theory is
explored. It is shown that
T--duality is always compatible with supersymmetry and simply changes a local
realization to a non--local one and vice versa.
Non--local realizations become natural
using classical parafermions of the underlying conformal field theory.
Examples presented include hyper--kahler metrics and the backgrounds for
the $SU(2) \otimes U(1)$ and $SU(2)/U(1) \otimes U(1) \otimes U(1)$
exact conformal field theories.

\footline{}

\vfill\eject

\footline={\hss\tenrm\folio\hss}% 	restores pagenumbers
\pageno=1
\vsize=46pc

%%%%%%%%%%%%%%%%%%%%%%%%%%%%%%%%%%%%%%%%%%%%%%%%%%%%%%%%%%%%%%%%%%%

\newsec{ Introduction }

This note is based on a lecture given in the Leuven conference
on Gauge Theories, Applied Supersymmetry and Quantum Gravity and contains
some of our recent work on Target space duality (T--duality) and supersymmetry
\refs{\basfep,\sferesto}. Further details can be
found in there and other papers that are directly
related to the subject \refs{\basusy\BKO\hassand-\AAB}.
Special care was taken so that the note is as introductory and self contained
as possible.

The description of certain phenomena in terms of effective field theories
might lead to paradoxes. One such case is the apparent incompatibility
between T--duality and supersymmetry which was first noticed in \basusy\
and subsequently with other examples in \BKO. These obstructions were observed
completely at the classical level for the case of extended world--sheet
supersymmetry and at 1--loop level in $\a'$ for target space supersymmetry.
T--duality is a stringy property and provides an equivalence between strings
propagating in different backgrounds \BUSCHER\ and as such it
should not lead to a real breaking of other genuine symmetries such as
supersymmetry. The natural explanation of this paradox is that,
non--local world--sheet effects associated with the
T--duality transformation replace a local realization of supersymmetry with a
non--local one, but never break supersymmetry which is always manifest at
the conformal field theory (CFT) level \basfep.
This point of view was also advocated in \refs{\hassand,\AAB}.

This issue is relevant for string phenomenology since if duality breaks or,
restores manifest supersymmetry \refs{\AAB,\sferesto} (we will use the terms
manifest and locally realized supersymmetry in an equivalent way), then this
phenomenon should be incorporated in supersymmetry breaking scenarios.
``Apparently'' non--supersymmetric backgrounds can qualify as vacuum solutions
to superstring theory when it is possible to restore manifest supersymmetry
through non--local world--sheet effects at the string level.
In addition,
it raises the possibility that various solutions that are of interest in
black hole physics or cosmology might have hidden supersymmetries in a string
context. Since this necessarily involves non--local world--sheet effects
it is important in our effort to understand the way string theory
could resolve fundamental problems in physics.

The rest of this section is devoted to the formulation of abelian T--duality
as a canonical transformation. In section 2 extended world--sheet
supersymmetry and its relation to T--duality is explored.
As examples we consider
hyper--kahler manifolds and the semi--wormhole exact solution of 4--dim string
theory. In the later case direct contact with the underlying CFT theory and
classical parafermions is made.
We end this note with section 3 where some related topics are briefly
discussed.

\vskip .1 cm
\no
$\underline {\rm T-Duality\ and\ canonical\ transformations }$:
The mechanism by which abelian T--duality changes local realizations of
supersymmetry to non--local ones is most transparent in its formulation
as a canonical transformation \AALcan. Here we follow a slightly different
route which especially suits our purposes.
The classical propagation of strings in a general target space with
metric $G_{\mu \nu}(X)$ and antisymmetric tensor field $B_{\mu \nu}(X)$ is
described by the 2--dim $\sigma$--model Lagrangian density
$\cL = Q^+_{\mu \nu} \del_+ X^{\mu} \del_- X^{\nu}$, where
%\eqn\boso{\eqalign{
%& S(x) = \ha \int Q^+_{\m\n} \del_+ X^\m \del_- X^\n \ ,\quad \m,\n = 0,1,
%\dots ,d-1 \ ,\cr}}
$Q^\pm_{\m\n}\equiv G_{\m\n} \pm B_{\m\n}$. The natural time coordinate on
the world--sheet is $\tau= \s^+ + \s^- $. Also will use $\s= \s^+ - \s^-$
to denote the position along the string
for fixed $\tau$. We assume there exist a Killing symmetry associated
with the vector field $\del/{\del X^0}$ and we denote the rest of the
coordinates by $X^i$, $i=1,\dots , d-1$. For convenience we will work in the
adapted coordinate system where the background fields are independent of $X^0$.
The canonical approach \AALcan\ starts by computing the conjugate
momenta to $X^\m$, $ P_\m = { \del \cL \ov \del \dot X^\m}$ (the dot denotes
derivative with respect to $\tau$)
\eqn\mome{ P_\m = \ha Q^+_{\m\n} \del_- X^\n + \ha Q^-_{\m\n} \del_+ X^\n \ .}
The transformations
%\eqn\tracan{ P_0 =  \del_\s \tilde X^0 \ ,\quad \del_\s X^0 = \tilde P_0 \ ,}
$ P_0 =  \del_\s \tilde X^0$, $\del_\s X^0 = \tilde P_0$
and the redefinitions $X^i = \tilde X^i$, $P_i = \tilde P_i$
preserve the Poisson bracket
%\eqn\poison{ [X^\m(\tau,\s),P_\n(\tau,\s')]= i \d^\m{}_\n\ \d(\s-\s')\ ,}
$ \{X^\m(\tau,\s),P_\n(\tau,\s')\}=  \d^\m{}_\n \d(\s-\s')$
and therefore constitute a canonical transformation.
The world--sheet derivatives
\eqn\woldev{ \del_\pm X^0\ \equiv\ \dot X^0 \ \pm\  \del_\s X^0 \ =\
 G_{00}^{-1}
(P_0 - \ha Q^+_{i0} \del_+ X^i - \ha Q^-_{i0} \del_- X^i)\ \pm\ \del_\s X^0\ ,}
transform as well.
After some algebra and using the analogous to \mome\
expression for the conjugate momentum to $\tilde X^0$, $\tilde P_0$, we obtain
\eqn\wores{\eqalign{ \del_\pm X^0 = &
\pm \ha (\tilde G_{00} + G_{00}^{-1})\del_\pm \tilde X^0
\ \pm\ \ha (\tilde G_{00} - G_{00}^{-1})\del_\mp \tilde X^0 \cr
& \pm \ha (\tilde Q^\pm_{i0} \mp G_{00}^{-1} Q^\pm_{i0})\del_\pm X^i
\ \pm\ \ha (\tilde Q^\pm_{0i} \mp G_{00}^{-1} Q^\pm_{0i})\del_\mp X^i \ .\cr}}
In order to preserve 2-dimensional Lorentz invariance we should have
world--sheet derivatives of the same chirality in both sides of \wores.
Setting to zero the relevant terms we relate the background fields in the
transformed model to those in the original one
\eqn\dualbou{\tilde G_{00} = G_{00}^{-1}\ ,\quad \tilde Q^\pm_{0i} =
\pm G_{00}^{-1} Q^\pm_{0i}\ ,\quad \tilde Q^+_{ij}= Q^+_{ij} -G_{00}^{-1}
Q^+_{i0} Q^+_{0j}\ .}
These are nothing but Buscher's duality transformation rules \BUSCHER.
Actually, the transformation of $Q^+_{ij}$
follows by identifying the coefficients of world--sheet derivatives in
$P_i=\tilde P_i$ after 2-dimensional Lorentz invariance was taken into
account. The conformal invariance also requires that the
corresponding dilaton field $\Phi$ is shifted by $\ln G_{00}$ \BUSCHER.
The transformed world--sheet derivatives under duality then become
\eqn\woderd{ \del_\pm X^0= \pm G_{00}^{-1} ( \del_\pm \tilde X^0 \mp
Q^\pm_{i0} \del_\pm X^i )= \pm \tilde Q^\pm_{\m 0} \del_\pm \tilde X^\m \ .}
This transformation amounts to a non--local redefinition of the
target space variable associated with the Killing symmetry,
\eqn\rednonl{ X^0=\int \tilde Q^+_{\m 0} \del_+ \tilde X^\m d \s^+
- \tilde Q^-_{\m 0} \del_- \tilde X^\m d\s^- \ .}
Despite the non--localities, the dual target space fields \dualbou\ are
locally related to the original ones. However, other geometrical
objects in the target space, are not bound to be always local in the
dual picture \basfep. This will be discussed extensively in the next section.

\vskip .1 true cm
\no
$\underline {\rm N=1\ world-sheet\ supersymmetry\ under\ duality}$:
It is well known that any background can be made $N=1$ supersymmetric \FRTO.
Thus
we do not expect any clash with duality (abelian and non--abelian T--duality
as well as S--duality) in this case. In fact one can
formulate abelian duality in a manifestly supersymmetric way by using $N=1$
superfields \hassand. The result is that the
transformation of $\del_\pm X^0$ is given by \woderd\ plus a quadratic term
in the fermions and that the fermions themselves transform as the world--sheet
derivatives in \woderd.
In other words bosons in the dual model are composites of bosons
and fermions of the original model.
This boson--fermion symphysis was first observed in
\zasusydu\ for the supersymmetric extension of the Chiral Model on $O(4)$ and
its non--abelian dual. For convenience we will not subsequently write
the dependence on the world--sheet fermions since, on general grounds,
it is completely dictated by the purely bosonic term \sferesto.

\newsec{ T--Duality and extended world--sheet supersymmetry }

In contrast with $N=1$,
it is well known that extended $N=2$ supersymmetry \refs{\zumino,\ALFR,\GHR}
requires that the background is such that an (almost) complex (hermitian)
structure $F^\pm_{\m\n}$ in each sector, associated to the right and
left-handed fermions, exists.
%The conditions to be satisfied are
%\eqn\comcond{ (F^\pm)^\m{}_\l (F^\pm)^\l{}_\n = -\d^\m{}_\n \ ,\quad
%F^\pm_{\m\n} +  F^\pm_{\n\m} = 0\ ,\quad D^\pm_\m (F^\pm)^\l{}_\r = 0\ ,}
%where $F^\pm_{\m\n}\equiv G_{\m\l} (F^\pm)^\l{}_\n$ and
%the generalized connections that are used to define the covariant derivatives
%include the torsion $H_{\m\n\r}=3 \del_{[\r} B_{\m\n]}$.
Similarly, $N=4$ extended supersymmetry \refs{\ALFR,\GHR,\PNBW}
requires that, in each sector,
there exist three complex structures  $(F_I^\pm)_{\m\n}$, $I=1,2,3$.
The complex structures are covariantly constant, with respect to generalized
connections that include the torsion, antisymmetric matrices and
in the case of $N=4$ they obey the $SU(2)$ Clifford algebra.
These conditions put severe restrictions on the backgrounds that admit a
solution. For instance in the absence of torsion the metric should be Kahler
for $N=2$ and hyper--kahler for $N=4$ \ALFR.

We are interested in finding the duality transformation properties of the
complex structures. We simply examine the 2--forms defined in each chiral
sector separately \basfep
\eqn\formch{  F^\pm_I = (F^\pm_I)_{\m\n} d X^\m \wedge d X^\n =
2 (F^\pm_I)_{0i} d X^0 \wedge d X^i \ + \
(F^\pm_I)_{ij} d X^i \wedge d X^j \ .}
%The $F^\pm_I$ are associated to right or left--handed fermions and
In order to find the correct transformation
properties under T--duality, we simply have to use the replacement
$d X^{\mu} \to \del_+ X^{\mu}$ for $F^+_I$ and
$d X^{\mu} \to \del_- X^{\mu}$ for $F^-_I$.
This is only meant to be a prescription for
extracting the relevant part of the complex structures under the
duality transformation \woderd. Then the dual complex structures in component
form are
\eqn\resdcom{ (\tilde F^\pm_I)_{0i}= \pm G_{00}^{-1} (F^\pm_I)_{0i}\ ,
\qq (\tilde F^\pm_I)_{ij}= (F^\pm_I)_{ij} \ + \ G_{00}^{-1}
\bl( (F^\pm_I)_{0i} Q^\pm_{j0} - (F^\pm_I)_{0j} Q^\pm_{i0}  \br)\ .}
There was a crucial assumption made implicitly, namely that in the adapted
coordinate system the complex structures
$F^\pm_I$ were independent of the Killing coordinate $X^0$ similarly to
background fields. It turns out that this is a correct assumption for the case
of $N=2$ extended world--sheet supersymmetry. However, it is not always true
in the case of $N=4$ extended supersymmetry where in certain cases two of the
complex structures depend on $X^0$ explicitly and they form an
$SO(2)$ doublet. Then, under duality they become
non--local due to the mechanism associated with \rednonl. A useful criterion
for when abelian T--duality preserves manifest $N=4$ extended world--sheet
(and target space) supersymmetry is \sferesto\foot{ It has
been used to prove that a marginal deformation by current--bilinears in the
Cartan subalgebra of a WZW model for a general quaternionic group
always breaks their manifest $N=4$ world--sheet supersymmetry \sferesto.}
\eqn\critii{ \del_\m Q^\mp_{0\n} (F^\pm_I)^{\m\n} = 0\ ,\quad I=1,2,3\ .}
In the example that we consider next, in which local $N=4$ is preserved under
duality, one can explicitly show that \critii\ is indeed true for all three
complex structures. In contrast, in the rest of the examples it is violated
for $I=3$, in agreement with the fact that T--duality in these cases destroys
manifest $N=4$, which then is realized non--locally.

\subsec{ Hyper--Kahler manifolds and T--duality }

Let us first consider 4-dim pure
gravitational backgrounds with $N=4$ extended supersymmetry, which are known
to be hyper--kahler self--dual manifolds,
that in addition have one Killing symmetry.
A complete classification of them exists and depends on whether or not the
covariant derivative of the corresponding Killing vector is self--dual \BOFI.
Accordingly, the Killing vector is of the translational or the
rotational type.

\vskip .1 cm
\no
$\underline {\rm The\ translational\ case}$:

In the case of 4--dim hyper--Kahler manifolds with a translational symmetry the
metric assumes the form \HAGITO
\eqn\tranmet{ ds^2 = V(dT + \Om_i dx^i)^2 + V^{-1} dX_i dX^i\ ,}
where $T\equiv X^0$ is the coordinate adapted for the translational Killing
vector field $\partial / {\partial T}$. Moreover, ${\Om}_{i}$ are
constrained to satisfy the special conditions
\eqn\selcon{\del_{i} V^{-1} =  {\epsilon}_{ijk} {\partial}_{j}{\Om}_{k} \ ,}
as a result of the self--dual character of the metric.
It also follows that $V^{-1}$ satisfies the 3--dim flat space
Laplace equation. Localized solutions of this equation correspond to
the familiar series of multi--centre Eguchi--Hanson gravitational
instantons or to the multi--Taub--NUT family, depending on the asymptotic
conditions on $V^{-1}$ (see, for instance, \EGHgra\ and references therein).
The three independent complex structures are \GIRU
\eqn\trancomp{ F^i= (dT + \Om_j dX^j)\wedge dX^i  -  \ha V^{-1} \e^{ijk}
dX^j \wedge dX^k\ .}
We see that the complex structures are invariant under constant shifts of $T$.
Thus under duality $N=4$ supersymmetry will remain locally realized.
Indeed the dual to \tranmet\ background is found using \dualbou\ to be
\eqn\dualtr{\eqalign{& d\tilde s^2= V^{-1} ( d\tilde T^2 + dX_i dX^i) \ ,\cr
& \tilde B = 2\ \Om_i d\tilde T \wedge dX^i\ ,\quad
\tilde{\Phi} = \ln V\ , \cr }}
%There is also a dilaton field $
where we have denoted by $\tilde T$ the Killing coordinate in the dual model.
For the dual complex structures we use \resdcom\ and find the result \basfep
\eqn\dualcomp{ \tilde F^i= V^{-1} (\pm d\tilde T\wedge dX^i
- \ha \e^{ijk} dX^j \wedge dX^k)\ ,}
which define a local realization of the $N=4$ world--sheet supersymmetry
for \dualtr.

\vskip .1 cm
\no
$\underline {\rm The\ rotational \ case}$:

In the case of 4--dim hyper--Kahler manifolds with a rotational
Killing symmetry, there exists a coordinate system ($\tau$, $x$, $y$,
$z$) in which the corresponding line element assumes the form
\eqn\rotamet{ds^2 = v(d \tau + {\omega}_{1} dx + {\omega}_{2} dy)^{2} +
v^{-1} (e^{\Psi} {dx}^{2} + e^{\Psi} {dy}^{2} + {dz}^{2}) \ .}
In these adapted coordinates the rotational Killing vector field is
$\partial / {\partial \tau}$ and all the components of the metric are
expressed in terms of a single scalar field $\Psi (x, y, z)$ \BOFI, so that
%\eqn\rotaexp{ v^{-1} = {\partial}_{z} \Psi\ ,\qq {\omega}_{1} = -
%{\partial}_{y} \Psi\ ,\qq{\omega}_{2} = + {\partial}_{x} \Psi\  ,}
$v^{-1} = \del_z \Psi$, $\om_1 = - \del_y \Psi$, $\om_2 =+ \del_x \Psi$ and
where $\Psi (x, y, z)$ satisfies the continual Toda equation
\eqn\rotato{(\del_x^2 + \del_y^2) \Psi + \del_z^2 e^{\Psi} = 0 \ .}
Examples of metrics which can be put into the form \rotamet\ include the
Eguchi--Hanson instanton and the Taub--NUT and
Atiyah--Hitchin metrics on the moduli space of the $SU(2)$ 2--monopole
solutions in the BPS limit.
In fact, the first two metrics exhibit both kinds of isometries, translational
and rotational (see for instance \GIRU).

Metrics with rotational Killing symmetry differ from those with
translational symmetry
in that not all three independent complex structures can be chosen to
be $\tau$--shift invariant. In fact, only one complex structure can be
chosen to
be an $SO(2)$ singlet, while the other two necessarily form an $SO(2)$ doublet.
We have explicitly \basfep
\eqn\rotasi{F_{3} = (d \tau + {\omega}_{1} dx + {\omega}_{2} dy) \wedge dz +
v^{-1} e^{\Psi} dx \wedge dy \ ,}
for the singlet and
\eqn\rotado{\pmatrix{F_{1} \cr F_{2}}= e^{{1 \over 2} \Psi}
\pmatrix{ \cos {\tau \over 2} & \sin {\tau \over 2} \cr
\sin {\tau \over 2} & - \cos {\tau \over 2}} \pmatrix{ f_1 \cr f_2}\ ,}
for the doublet, where
\eqn\rotaff{f_{1} = (d \tau + {\omega}_{2} dy) \wedge dx
- v^{-1} dz \wedge dy \ ,\qq
f_{2} = (d \tau + {\omega}_{1} dx) \wedge dy + v^{-1} dz \wedge dx \ .}
The T--duality transformation with respect to $X^{0} \equiv \tau$,
yields the background
\eqn\rotdu{\eqalign{&
d \tilde s^2 = v^{-1} (e^{\Psi} d x^{2} + e^{\Psi} d y^{2} +
d z^{2} + d {\tilde \tau}^{2}) \ ,\cr
&\tilde B = 2\ d \tilde \tau \wedge ({\om}_{1} dx + {\om}_{2} dy)\ ,\quad
\tilde{\Phi} = \ln v\ .\cr } }
%and the dilaton field $
The complex structure \rotasi\ will
remain local in the dual picture, assuming the form
\eqn\dualrtsi{\tilde F_3^{\pm} = v^{-1} (\pm d \tilde \tau \wedge dz + e^{\Psi}
dx \wedge dy) \ .}
In contrast, the complex structures forming the doublet \rotado\ become
non--local and are given by
\eqn\rduado{\pmatrix{\tilde F_{1} \cr \tilde F_{2}}= e^{{1 \over 2} \Psi}
\pmatrix{ \cos {\tau \over 2} & \sin {\tau \over 2} \cr
\sin {\tau \over 2} & - \cos {\tau \over 2}} \pmatrix{\tilde f_1 \cr \tilde
f_2}\ ,}
where the 2--forms dual to \rotaff\ are
\eqn\rtff{\tilde f_1^{\pm} = v^{-1} (\pm d \tilde \tau \wedge dx - dz
\wedge dy )\ ,\qq \tilde f_2^{\pm} = v^{-1} (\pm d \tilde \tau \wedge dy
+ dz \wedge dx )\ .}
The non--localities are due to the explicit dependence on
$\tau $ which is not a variable in the dual model but rather a functional
of the dual variables
\eqn\taaau{\tau = \int (v^{-1} \partial_+ \tilde \tau -
 {\omega}_{1} \partial_+ x
- {\omega}_{2} \partial_+ y) d\s^+ - (v^{-1} {\partial_-} \tilde \tau +
{\omega}_{1} {\partial_-} x + {\omega}_{2} {\partial_-} y)
d \s^-\ .}
Moreover, $\tilde F_1^{\pm}$ and $\tilde F_2^{\pm}$ are not
covariantly constant on--shell \refs{\basfep,\hassand}.
Nevertheless, they can still be used to define a supersymmetry.
We found that the local realization of $N=4$ world--sheet
supersymmetry breaks down to $N=2$, with ${\tilde{F}}_{3}^{\pm}$ providing the
relevant pair of complex structures.
At the level of the corresponding superconformal field
theory, part of the $N=4$ world--sheet supersymmetry will be realized
with operators that have parafermionic (non--local) behavior.
We do not have an exact conformal field theory description
of the string gravitational background \rotamet, in order to
illustrate this point in all generality. For this reason we will
examine the question in the special case of a 4--dim semi--wormhole
solution and its rotational dual background, where an exact
description is available in terms of the $SU(2)\otimes U(1)$ WZW model and its
derivatives.
%We will see later that the parafermion currents of
%the $SU(2)/U(1)$ coset model describe the non--local structure of the
%dual 2--forms ${\tilde{F}}_{1}^{\pm}$, ${\tilde{F}}_{2}^{\pm}$.
%thus providing the explicit construction of a non--locally realized
% $N=4$ superconformal algebra with $\hat{c} = 4$ \KAFK.

\subsec{ Complex structures and parafermions }

A semi--wormhole solution of 4--dim string theory provides
an exact conformal field theory background with $N=4$
world--sheet supersymmetry \refs{\CHS,\KAFK}. The $N=4$ superconformal
algebra can be locally realized in terms of four bosonic currents, three
non--Abelian $SU(2)_{k}$ currents and one Abelian current with
background
charge $Q = \sqrt{2/(k+2)}$, so that the central charge is $\hat{c} =
4$. There are also four free--fermion superpartners and the solution is
described by the $SU(2)_{k} \otimes U(1)_{Q}$ supersymmetric WZW
model.\foot{The realization of the $N=4$ superconformal algebra in
terms of $SU(2)$ currents was first considered in \STP.}
The background fields of this model are given by
\eqn\bacworm{\eqalign {
&d s^2 = d\r^2 + d\vphi^2 + \sin^2\vphi\ d\psi^2 + \cos^2\vphi\ d\tau^2 \ ,\cr
& B_{\tau\psi}=  \cos^2\vphi \ ,\quad \Phi = 2 \r \ .\cr } }
The analysis that follows was essentially performed in \basfep\ in a
slightly different parametrization.
Among the three complex structures one is an $SO(2)$ singlet
\eqn\siworm{
 F^\pm_3 = d\r \wedge (\cos^2\vphi \ d\tau\ \pm\ \sin^2\vphi\ d \psi) \
+\ \ha \sin 2\vphi ( d\tau \ \mp\  d \psi) \wedge d \vphi \ ,}
and the other two form an $SO(2)$ doublet
\eqn\comdd{ \pmatrix{ F^\pm_1 \cr  F^\pm_2}=
\pmatrix{ \cos(\tau \pm \psi)
& \sin(\tau \pm \psi) \cr -\sin(\tau\pm \psi)
& \cos(\tau \pm \psi)\cr}\pmatrix{f^\pm_1 \cr f^\pm_2} \ ,}
with the definitions
\eqn\tform{\eqalign{ & f^\pm_1= - d\r\wedge d\vphi\ \pm \ \ha
\sin 2\vphi\ d\tau \wedge d\psi \cr
& f^\pm_2= - \ha \sin 2\vphi \  d\r \wedge ( d\tau
\ \mp \ d \psi) \ + \  (\cos^2 \vphi\ d\tau\ \pm\ \sin^2\vphi\ d \psi)\wedge
d\vphi  \ .\cr } }
T--duality corresponding to the Killing vector
$\del/{\del \psi}$ gives the background \RSS
\eqn\dualworm{\eqalign{
& d\tilde s^2= d\vphi^2 + \cot^2\vphi\ d\a^2 + d\b^2 + d\r^2 \ ,\cr
& \tilde \Phi = 2\r + \ln(\sin^2\vphi )\ ,\cr }}
with zero antisymmetric tensor,
which corresponds to the $SU(2)_k/U(1) \otimes U(1) \otimes U(1)_Q$ model,
and where the redefinitions $\a=\tilde \psi -{\tau\ov 2}$ and
$\b=\tilde \psi +{\tau\ov 2}$ have been made. Under the T--duality the dual
complex structure to $F_3^\pm$ becomes
\eqn\dulftt{ \tilde F_3= d\r \wedge d\b + \cot\vphi \ d\vphi \wedge d\a\ ,}
and defines a local supersymmetry. Since there is no torsion there is
no distinction between the + and the -- components.
Similarly the 2--forms \tform\ become
\eqn\dualff{\tilde f_1 = - d\r \wedge d\vphi - \cot\vphi\ d\a \wedge d\b\ ,\qq
\tilde f_2= \cot\vphi\ d\r \wedge d\a + d\b \wedge d\vphi \ .}
However, the dual complex structures become non--local due to the explicit
appearance of $\psi$ in \comdd.
Using the non--local relation to the dual model variables
\eqn\duava{ \psi = \int (\cot^2\vphi \del_+ \a + \del_+ \b )d\s^+  -
(\cot^2\vphi \del_-\a + \del_- \b ) d\s^- \ ,}
we write the dual complex structures in the suggestive forms \basfep\foot{
Notice that,
there is a distinction between + and the -- components below even though
the torsion vanishes. This is a novel characteristic of non--local
realizations of supersymmetry \sferesto.}
\eqn\comdu{\eqalign{&\tilde F^+_1 = \Psi_+ \wedge (d\r + i \b) + \Psi_- \wedge
 (d\r - i d\b) ,\      % \quad
\tilde F^+_2= i \Psi_+ \wedge (d\r + i \b) - i \Psi_- \wedge
 (d\r - i d\b)\ ,\cr
 &\tilde F^-_1 = \bar \Psi_+ \wedge (d\r - i \b) + \bar\Psi_- \wedge
 (d\r + i d\b), \ % \quad
\tilde F^-_2=- i \bar\Psi_+ \wedge (d\r - i \b) + i \bar\Psi_- \wedge
 (d\r + i d\b)\ ,\cr }}
where the parafermionic-type 1--forms are defined as
\eqn\paraff{ \Psi_\pm=(d\vphi \pm i \cot\vphi \ d\a) e^{\pm i(\b-\a + \psi)}\ ,
\qq \bar\Psi_\pm = (d\vphi \mp i \cot\vphi \ d\a) e^{\pm i(\a-\b + \psi)}\ ,}
and are non--local due to \duava.
They have a natural decomposition in terms of $(1,0)$ and $(0,1)$ forms on the
string world--sheet
\eqn\decof{ \Psi_\pm =\Psi^{(1,0)}_\pm d\s^+ \ + \
\Psi^{(0,1)}_\pm d\s^- \ ,\quad
\bar\Psi_\pm = \bar\Psi^{(1,0)}_\pm d\s^+ \ + \
\bar\Psi^{(0,1)}_\pm d\s^- \ .}
It can be easily verified using the classical equations of
motion for the model \dualworm\ that the chiral and anti--chiral conservation
laws
\eqn\onss{ \del_- \Psi^{(1,0)}_\pm=0\ ,\qq \del_+ \bar\Psi^{(0,1)}_\pm=0 \ ,}
are obeyed. In fact in this case $\Psi^{(1,0)}_\pm$ and
$\bar \Psi^{(0,1)}_\pm$
are nothing but the classical parafermions for the $SU(2)_k/U(1)$ coset with
the field $\b$ actually providing the dressing to the full 4--dim model
$SU(2)_k/U(1) \otimes U(1) \otimes U(1)_Q$. Thus the original local $N=4$
world--sheet supersymmetry breaks to a local part corresponding to \dulftt,
and the rest is realized non--locally using the non--local complex structures
\comdu. At the (super)CFT level this is manifested by a
replacement of the three $SU(2)_k$ currents by two $SU(2)_k/U(1)$ parafermions
and one free boson in the realization of the $N=4$ superconformal
algebra \KAFK.

\newsec{ Further developments and comments }

In this section we briefly mention related topics that we had no space
to extensively analyze in the main text.

\vskip .1 true cm
\no
$\underline {\rm T-duality\ and\ target\ space\ supersymmetry }$:
A similar clash between T--duality and supersymmetry on the target space takes
place as well \basusy.
The conventional definition of a background with unbroken target space
supersymmetry requires that solutions to the Killing spinor equations exist.
In the presence of rotational--type Killing vectors abelian T--duality breaks
manifest target space supersymmetry in the sense that no Killing spinors exist
in the dual background \refs{\basusy,\BKO}.
The crucial difference with the extended
world--sheet supersymmetry case is that here the effective field
theory is not enough at all to understand the fate of target space
supersymmetry under duality, which neverthelss does not destroy the
supersymmetry in a string setting.
Finding non--local Killing spinors \hassand\ is not adequate to generate
the whole supersymmetry transformation.
An approach to the problem using CFT concepts has been made in \AAB, where
again the use of parafermions becomes necessary.
Finally, let us mention that breaking of manifest target space supersymmetry
occurs hand and hand with breaking of local $N=4$ extended
world--sheet supersymmetry. However, although
in the latter case the $N=2$ part remains local,
the local target space supersymmetry completely breaks.
This is atributed to the
relation of Killing spinors and complex structures \kilcom,
i.e, $F_{\m\n} = \bar \eta \G_{\m\n} \eta $, which makes possible to construct
a local complex structure out of a non--local Killing spinor.

\vskip .1 true cm
\no
$\underline {\rm Theorems\ revised  }$:
We have seen that, in a string theoretical
setting, non--local complex structures are equally acceptable as local ones,
since they capture stringy effects that are manifest at the (super)CFT level.
Hence, it is important to investigate the conditions under which
there is non--locally
realized extended supersymmetry in general, without preassuming its origin.
This was done in \sferesto. In addition, all theorems that were
proved in the past in the context of 2--dim supersymmetric $\s$--models
with extended supersymmetry (see for instance \PNBW\ and references therein),
always assumming local realizations,
need to be revised, since they clearly fail in the presence of non--local
realizations. For instance, $N=4$ and zero torsion does not imply Ricci
flatness (the reader can easily check that by computing the Ricci tensor
for the simple metric in \dualworm). Steps towards this direction have been
made \sferesto.

\vskip .1 true cm
\no
$\underline {\rm Non-abelian\ T-duality\ and\ supersymmetry }$:
Let us just consider the effect of the
non--Abelian duality transformations on pure gravitational backgrounds with
extended $N=4$ world--sheet supersymmetry.
For $SO(3)$--invariant metrics, the complex structures either can be
$SO(3)$ singlets, thus remaining invariant under the non--Abelian
group action, $ \cL_J (F_I)_{\mu\nu}=0$, or form an $SO(3)$ triplet when
$\cL_J (F_I)_{\mu\nu} = {\e}_{JIK} (F_K)_{\mu\nu} $.
The Eguchi--Hanson metric corresponds
to the first case, while the Taub--NUT and the Atiyah--Hitchin
metrics to the second \GIRU.
It should be clear then that the dual version of the Eguchi--Hanson instanton
with respect to $SO(3)$
will have an $N=4$ world--sheet supersymmetry locally realized.
On the other hand, performing the non--Abelian $SO(3)$--duality
to the Taub--NUT and the Atiyah--Hitchin metrics will result in a total loss
of all the locally realized extended world--sheet supersymmetries \basfep.
We expect to
have a non--local realization of supersymmetry in such cases with three
non--local complex structures that satisfy the general conditions of \sferesto.
There is now some work on non--Abelian
duality and canonical transformations in the target space
\zacdual\ which should be helpful in uncovering the hidden
non--local supersymmetry these models are expected to have.

\bs
\centerline{ \bf Acknowledgments }

I would like to thank the organizers for their warm hospitality during the
conference.

\listrefs
\end